\documentclass[preprint,aps]{revtex4}
\usepackage{graphicx}
\usepackage{dcolumn}
\usepackage{bm}
 
\newcommand{\be}{\begin{equation}}
\newcommand{\ee}{\end{equation}}

\newcommand{\eq}[1]{Eq. (\ref{#1})} 
 
\newcommand{\Fig}[1]{Fig. (\ref{#1})} 
 \newcommand{\eqa}{\begin{eqnarray}}
\newcommand{\eeq}{\end{eqnarray}}  
\begin{document}
\title{Turbulence and transport in two-dimensional magnetized electron plasmas}
\vspace{3cm}

\author{Dastgeer Shaikh} 
\email{dastgeer@ucr.edu}
\author{P. K. Shukla}
\email{ps@tp4.rub.de}
\affiliation{Institute of Geophysics and Planetary Physics (IGPP),\\
University of California, Riverside, CA 92521. USA.\\
}
\affiliation{Institut f\"ur Theoretische Physik IV and Centre for
Plasma Science and Astrophysics, Fakult\"at f\"ur Physik und
Astronomie, Ruhr-Universit\"at Bochum, D-44780 Bochum, Germany}

\vspace{2.in}
\begin{abstract}
Electron plasmas confined by an external magnetic field exhibit
variations in a two-dimensional plane orthogonal to the confining
magnetic field. A nonlinear fluid simulation code to investigate 
the properties of 2-D electron plasma wave turbulence in a nonuniform
magnetoplasma  has been developed.  It is found that the presence of 
the density gradient convection by mean electric fields considerably 
influence the characteristic nonlinear interaction processes, such as the energy 
cascades and the cross-field electron transport. The initial random turbulent state evolves 
towards an intermittent state where forward cascade of vorticity coexists 
with an inverse cascade of electric potential fluctuations. The latter lead 
to the formation of large scale entities in 2D electron plasmas and can be 
alternatively understood by seeking exact nonlinear coherent vortex solutions 
in the form of a dipolar-like configuration.  The energy cascades are governed
typically by the Kolmogorov-like $k^{-5/3}$ spectrum.  In agreement
with the experimental observations, we find that the electron
transport is improved significantly by the application of an
externally imposed electric field.
\end{abstract}
\maketitle
\newpage

\section{Introduction}

Electron plasma columns are routinely confined in experimental devices
by means of an external magnetic field for hours. While such non-neutral 
plasmas \cite{davidson} exhibit excellent stability property under ideal conditions, 
there exist critical issues that can potentially destabilize the electron
plasma columns. For instance, the asymmetry of confining field and the
presence of neutral gas particles are reported to often cause damages
to the stability of the electron plasma columns \cite{driscol}. During
the process of rotation around the ambient field, the electron plasma
particles collide with each other resulting thus in a cross-field
electron transport. The cross field transport is therefore dominated by the
electron-electron collisions in an electron plasma column, whereas
collisions with the background neutrals do not contribute to the
transport.  However the advent of a rotating wall electric field
imposed externally on the electron plasma column leads to an
improvement in the stability \cite{driscol}. An application of the
externally applied radial electric field ${\bf E}$ in the presence of
the confining orthogonal magnetic field ${\bf B}$ imparts a net
poloidal drift ${\bf v}_\perp \approx (c/B^2){\bf E}\times{\bf B}$ on
the electron plasma column. Consequently, the plasma column rotates
around the external magnetic field which supposedly enhances its
stability.

There exist morphological correspondence between the guiding 
center flow of electrons in a strongly magnetized electron
plasma and the two-dimensional (2D) incompressible Euler 
fluid \cite{okuda}. In a uniform plasma, the electron fluid dynamics 
is governed by the Navier-Stokes equation (NSE), which basically describe
the evolution of the electron vorticity. The NSE equation admits 
interesting electron fluid turbulence behavior and localized 
vortex motions in the form of clumps (positive vorticity) and 
holes (negative vorticity) \cite{schecter}. The dynamics of electron
plasma vortices in background vorticity distribution has been 
investigated both analytically and numerically \cite{schecter} as
well as experimentally \cite{kiwamoto00, kiwamoto07}. 

In a nonuniform electron magnetoplasma containing equilibrium electron density 
gradient and fixed ion background, the dynamics of 2D electron drift modes (EDMs) 
is governed by the modified NSE \cite{shukla89}. Possible stationary solutions 
of the latter can be cast in the form of a global vortex pattern \cite{shukla89,sen} 
and a double vortex \cite{shukla89,shukla94,hall}. Vortex structures have been 
observed \cite{ker91,rose92,peu93} in low-temperature laboratory devices in which 
electron clouds are confined by dc electric and magnetic fields. Furthermore, 
large amplitude EDMs can parametrically excite electron zonal flows \cite{shukla06}, 
which play a decisive role in electron confinement in nonuniform magnetized plasmas.

In this paper, we present the turbulent properties of nonlinearly interacting 
low-frequency (in comparison with the electron gyrofrequency) electron plasma 
waves (or EDMs) in a nonuniform magnetoplasma containing the electron density inhomogeneity. 
The manuscript is organized in the following fashion.  In Sec. II, we discuss our 2D model 
equation for nonlinear EDMs and the possibility of frequency condensation due to the mode 
coupling process. Simulation results for a random initial turbulence state are described 
in Sec. III.  In Sec. IV, we seek to understand the emergence of large-scale
coherent structures by virtue of exact analytic solutions. The evolution
of turbulent energy and the corresponding transport are discussed
in Secs. VI and VII, respectively.  The conclusions are contained in
Sec. VIII.

\section{Basic Equations}

Let us consider a strongly magnetized electron plasma in the  presence of 
the equilibrium  density gradient ($\partial n_0/\partial x)$
across $\hat {\bf z} B_0$, where $\hat {\bf z}$ is the unit vector
along the $z$ direction in a Cartesian coordinate system and $B_0$ 
is the strength of the external magnetic field. The ions are 
supposed to form an immobile uniform neutralizing background.
The perpendicular component of the electron fluid velocity in the
presence of nonlinearly coupled low-frequency (in comparison 
with the electron gyrofrequency $\omega_{ce} =eB_0/m_ec$, where 
$e$ is the magnitude of the electron charge, $m_e$ is the electron
mass and $c$ is the speed of light in vacuum) electric field
${\bf E} = -\nabla \phi$, where $\phi$ is the electrostatic
potential, is
\be
{\bf u}_{e\perp} \approx \frac{c}{B_0}\hat {\bf z}
\times \nabla \phi + \frac{c}{B_0\omega_{ce}}\left(\frac{d}{dt}
+\nu_e -\mu_e \nabla_\perp^2\right)\nabla_\perp \phi,
\ee
where $d/dt =(\partial/\partial t) +(c/B_0)\hat {\bf z} \times \nabla \phi$,
$\mu_e$ is the electron collision frequency, and $\mu_e$ is the electron kinematic
gyro-thermal-viscosity. The nonlinearity in (1) comes from the nonlinear
electron polarization drift. The electron advection term $v_{ez}\partial/\partial z$
does not appear in $d/dt$ because of our 2D approximation. 

Inserting (1) into the electron continuity equation and using the Poisson
equation, we obtain the modified NSE
\be
\label{vor}
\left[ \frac{d}{dt} + \frac{\omega_{pe}^2}{\omega_{H}^2}
(\mu_e -\mu_e \nabla_\perp^2) \right] \nabla_\perp^2 \phi 
+ \frac{\omega_{pe}^2 \omega_{ce}}{\omega_H^2}K_n \frac{\partial \phi}{\partial y} 
= 0,
\ee 
where $\omega_{pe} =(4\pi n_0e^2/m_e)^{1/2}$ is the electron plasma frequency,
$\omega_H =(\omega_{pe}^2+\omega_{ce}^2)^{1/2}$ is the upper-hybrid resonance
frequency, and $K_n = -\partial n_0/\partial x$. In the absence of  
nonlinear interactions, Eq. (2) yields
\be
\omega = k_y u_0 +\frac{\omega_{pe}^2\omega_{ce}k_y K_n}{\omega_H^2 k_\perp^2}
-i \frac{\omega_{pe}^2}{\omega_{H}^2} \Gamma,
\ee
which is the frequency of the damped Doppler-shifted EDMs in the local approximation 
(viz.  the wavelength being much smaller than the density gradient scalelength $K_n^{-1}$),
and $\Gamma =\nu_e + \mu_e k_\perp^2$ is the damping rate.
Here $\omega$ is the wave frequency, ${\bf k} = {\bf k}_\perp \equiv (k_x,k_y)$ 
is the wave vector, $u_0 =(c/B_0)\partial \phi_0/\partial x \equiv -(c/B_0)E_0$ is the 
equilibrium electron drift in the presence of the dc electric field $E_0$, and 
$k_\perp^2 =k_x^2 + k_y^2$.  In the absence of the dc electric field and dissipation, 
we observe from (3) that there exists frequency condensation at $k_y \gg k_x$. On the 
other hand, in a uniform plasma without $E_0$, we have the electron convective cell 
mode \cite{okuda}, $\omega =-i (\omega_{pe}^2/\omega_{H}^2) (\nu_e + \mu_e k_\perp^2)$, 
which causes the cross-field electron transport due to random walk of the electrons in 
the electron convective cell electric field, even in a thermal equilibrium plasma \cite{okuda}. 

Equation (2) describes the evolution of the electron fluid vorticity in 2D 
plane, e.g. in the $(x,y)$ plane. The kinematic electron-thermal gyroviscosity 
(say due to turbulent fluctuations) has been introduced to accommodate the damping 
of fluctuations at short scales. 

To investigate the turbulence dynamics, we should have the knowledge of 
constant of motions of Eq. (2). The latter, without dissipation, admits two 
conserved quantities, namely the  energy and enstrophy (squared vorticity)  
\be
\label{en}
W =\int (\nabla_\perp \phi)^2 dxdy, ~~ \Omega= \int (\nabla_\perp^2\phi)^2 dxdy, 
\ee
which show that the energy and mean squared vorticity are conserved ideally (or inviscidly) 
by 2D nonlinear interactions in a strongly magnetized electron plasma.  

\section{simulation results -Formation of Coherent Structures}

The modified NSE \eq{vor} has been integrated numerically with the help 
of a fully de-aliased pseudospectral scheme.  Periodic boundary conditions 
are imposed along the $x$ and $y$-directions.  The electrostatic potential 
is discretized in a Fourier space using $f({\bf k},t)=\sum_{\bf k} f({\bf r},t) 
\exp(-i {\bf k}\cdot{\bf r})$.  All fluctuations in our simulations are
initialized with a Gaussian random number generator to ensure that the
Fourier components are all spatially uncorrelated and randomly phased.
This ensures that the choice of initial state is highly isotropic,
i.e.  $ k_x \approx k_y $ at $t=0$. Similarly, the boundary conditions
(periodic in $x,y$ directions) do not impose any kind of anisotropy.
Moreover, the results to be presented here are independent of the size
of computational domains, number of Fourier modes, as well as the
integration time steps.

\begin{figure}
\includegraphics[width=16.cm]{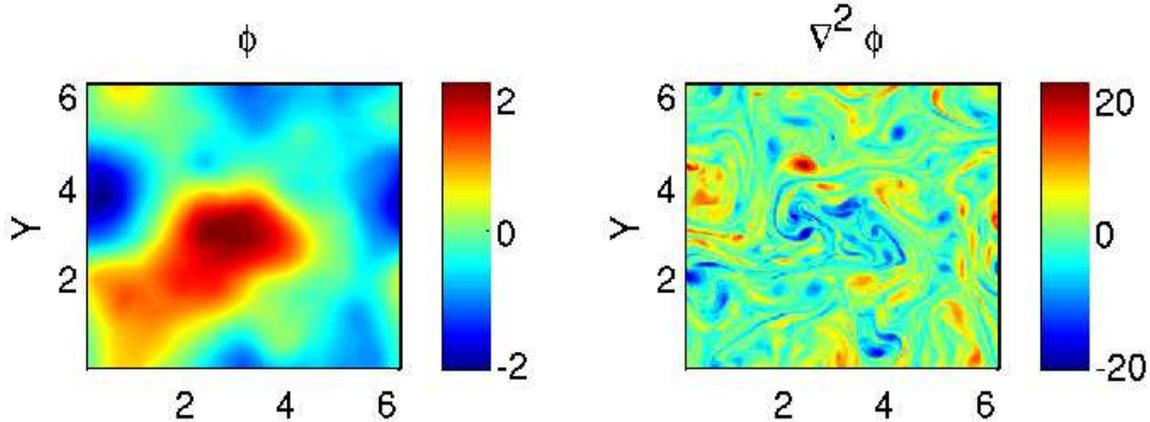}
\caption{\label{fig1} Simulation results from a random initial state
  lead to the formation of large scale coherent mode in potential
  fluctuations owing to an inverse cascade process (see left
  fig). Figure on the right shows the small scale eddies that are formed
  essentially from the forward cascades of electron fluid vorticity at
  a similar time. This is called a dual cascade phenomenon, observed
  ubiquitously  in many 2D turbulent systems. Here we have chosen
  $\Gamma/\omega_{ce} =0.01$ and $\beta =(\omega_{pe}^2/\omega_H^2)K_n L = 2$, 
  where $L$ is the box size.}
\end{figure}

The initially normalized energy spectrum, peaked at $ k_{\rm min}$, is
chosen to lie within the wavenumber interval $k_{\rm min} < k < k_{\rm max}/2$.  
During the evolution of our simulations, the turbulence
eventually evolves through vortex-merging in which like-signed smaller
length-scale fluctuations merge to form relatively large-scale
fluctuations. The process continues until all merging has occurred to
finally form the largest scale coherent vortex dominated by the
minimum allowed $k$ in the simulation.  The inertial range turbulent
cascades in such a manner leads to the formation of large scale
structures in potential fluctuations. By contrast, the vorticity
structures break up and progressively form smaller eddies by means of
a forward cascade processes. The large scale potential field
nonetheless coexist with the small scale vorticity fluctuations and
lead essentially to a dual cascade process.  This is shown in
\Fig{fig1} when the turbulence has reached its saturated state.  The
formation of large scale potential and small scale vorticity
structures through nonlinear interactions can be understood in the
context of a dual cascade phenomenon as the 2D electron plasma system
admits two inviscid invariants. Under this process the potential
cascades towards longer length-scales, while the fluid vorticity
transfers spectral power towards shorter length-scales.  The randomly
excited Fourier modes transfer the spectral energy in the inertial
range by conserving the constants of motion (i.e. the inviscid
invariants) in $k$ space. This leads to a statistically stationary
inertial regime associated with the forward and inverse cascades.  The
dual cascades in the electron plasma wave turbulence is similar to the one that
occurs in 2D hydrodynamic turbulence.  The energy
cascades towards smaller scales in the simulations is terminated
essentially by a kinematic gyro-viscous damping. The latter is efficient at the smaller
turbulent scales. The dual cascade processes reported here are further
consistent with the number of inviscid constants (see \eq{en}) as
admitted by the underlying system of \eq{vor}.

Note that the presence of the density gradient convection caused by mean 
electric field does not quantitatively alter the characteristic nonlinear 
features of 2D electron plasma wave turbulence, as it exhibits a cascade property 
similar to that of the Charney-Hasegawa-Mima equation.  The evolving relaxation 
of the turbulent fluid is independent of spatial and temporal resolutions, as
well as higher turbulent Reynolds numbers. The latter slow the rate of
relaxation, while the qualitative physics remains more or less
unaltered.  Interestingly, we notice the formation of thin current
sheet like structures in the evolution of the vorticity as shown in
\Fig{fig1}. Furthermore, the formation of the observed large scale
structures in our simulations can be understood from an exact nonlinear
analytic solution of \eq{vor}. This is described below.

\section{Analytical Solutions for A Double Vortex}

An analytic understanding of the emergence of coherent nonlinear
structures observed in our fluid simulation can be achieved merely
{\it qualitatively} by considering exact nonlinear solutions of the
potential in \eq{vor}.  The underlying physical mechanism for the 
formation of coherent structures rests on the nonlinear interactions
between different scalesize modes. In saturated states, 
the vector product nonlinearity overwhelms the dissipation, 
and a balance between the wave dispersion and nonlinearity 
gives rise to long-lived stable vortex structures that we 
observed in our simulations.

The two-dimensional traveling wave solutions of \eq{vor}, without dissipation, 
in the moving frame can be obtained by supposing that $x=x, \xi=y-ut, t=t$, 
where $u$ being the speed of the vortex. This further transforms the fluctuating
quantities into $\phi(x,y,t) = \phi(x,\xi)$. The potential fluctuation
obeys the localized boundary conditions $\phi \rightarrow 0$ as
$y\rightarrow \infty$ for all $x$. The condition of localization
associated with the coherent structures further enables us to use the
periodic boundary conditions in computational domain of the nonlinear
fluid simulations.  The perturbed entities, or coherent structures, in
the local region are far from the boundary and can translate
freely. Hence no boundary effects are considered in our
analysis. Boundary effects can nevertheless be important for certain
processes, but this issue is well beyond the scope of the present
work. Moreover, the $x-\xi$ plane in the simulations is orthogonal to
the background magnetic field.  On using the transformed co-ordinate
systems, the electron vorticity \eq{vor} can be translated into a
co-ordinate system that is moving along $\xi$-direction with the vortex
speed.  Following the previous treatment of the vortex theory 
\cite{larichev,horton,pokh,stenflo,horton1,shukla95}, we write (2) in 
the stationary frame 
\be
D_\xi\left(\nabla_\perp^2 \phi -\frac{u_*}{(u-u_0)}\phi\right) =0,
\ee
where $D_\xi =(\partial/\partial \xi)- [c/(u-u_0)B_0][(\partial \phi/\partial x)-
(\partial \phi/\partial \xi)\partial/\partial x]$, $\nabla_\perp^2 \phi 
=(\partial^2 \phi/\partial x^2) + (\partial^2\phi/\partial \xi^2$, and 
$u_* =-(\omega_{pe}^2\omega_{ce}/\omega_H^2) K_n$. Equation (5) is satisfied by 

\be
\nabla_\perp^2 = C_1 \phi + C_2 x,
\ee
where $C_1$ and $C_2$ are constants, and $C_1 +[(c/(u-u_0)B_0] C_2 =u_*/(u-u_0) =\alpha^2$. 
For $\alpha^2 > 0$, Eq. (6) admits a double vortex \cite{larichev}. 
The profiles of the latter in the outer ($r > R$, where $r^2 =x^2+ \xi^2$ and $R$ is 
the vortex radius) and inner ($r < R$; centered at the point $x=\xi=0$) regions are, 
respectively,

\be
\phi(r,\theta) = \phi_0 K_1(\alpha r)\cos \theta,
\ee
and

\be
\phi(r,\theta) =\left[\phi_i J_1(\gamma r) + D_i r/\gamma^2\right]\cos \theta,
\ee
where $\cos \theta = \xi/r$, $\phi_0 = D_i R/(\alpha^2 +\gamma^2)K_1(\alpha R)$, $\phi_i =
- \alpha^2 D_i R/\gamma^2(\alpha^2 +\gamma^2)J_1 (\alpha R)$, $D_i =(\alpha^2 +\gamma^2)(u-u_0)B_0/c$,
and the constant $\gamma$ has to be determined from the transcendental equation

\be
\frac{K_2(\alpha R)}{\alpha K_1(\alpha R)} =-\frac{J_2(\gamma R)}{\gamma J_1(\gamma R)}.
\ee
Here $J_1 (J_2)$ and $K_1 (K_2) $ are the Bessel and modified Bessel functions of the first (second) 
order, respectively. It turns out that the presence of the density gradient is essential
for the formation of a double vortex.

\begin{figure}
\includegraphics[width=10.cm]{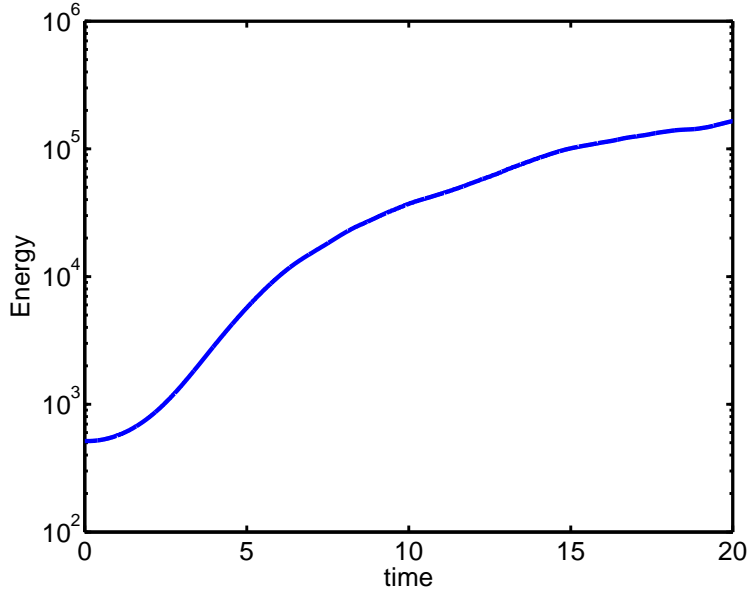}
\caption{\label{fig2}  Evolution of the total energy in the simulation. }
\end{figure}

\section{Evolution of Turbulent Energy and Spectrum}

The evolution of the turbulent energy associated with the potential
and vorticity fields is shown in \Fig{fig2} where the linear and
nonlinear phases of the evolution are clearly marked by the curve. The
nonlinear \eq{vor}, in the absence of dissipation, conserves the total
energy.  In the presence of dissipation, small scales are
dissipated. Since the 2D electron plasma system is dominated by the
smaller $k$'s, the large scale turbulent eddies tend to grow and
contain most of the turbulent energy. The volume integrated energy
therefore grows in time until the steady state is reached. The
evolution of the energy is shown in \Fig{fig2}.  This is further
consistent with the formation of large-scale coherent convective
cells/flows in which the energy associated with turbulence must evolve
in order to relax the turbulence into a well organized coherent
structure.  After nonlinear interactions are saturated, the energy
in the turbulence does not grow and remains nearly unchanged
throughout the simulations.  Correspondingly, the energy transfer rate
shows a significant growth during the linear and initial nonlinear
phases (not shown in \Fig{fig2}). However, when  nonlinear
interactions saturate, the nonlinear transfer of the energy in the
spectral space amongst various turbulent modes becomes inefficient and
the energy transfer per unit time tends to become negligibly small.

\begin{figure}
\includegraphics[width=10cm]{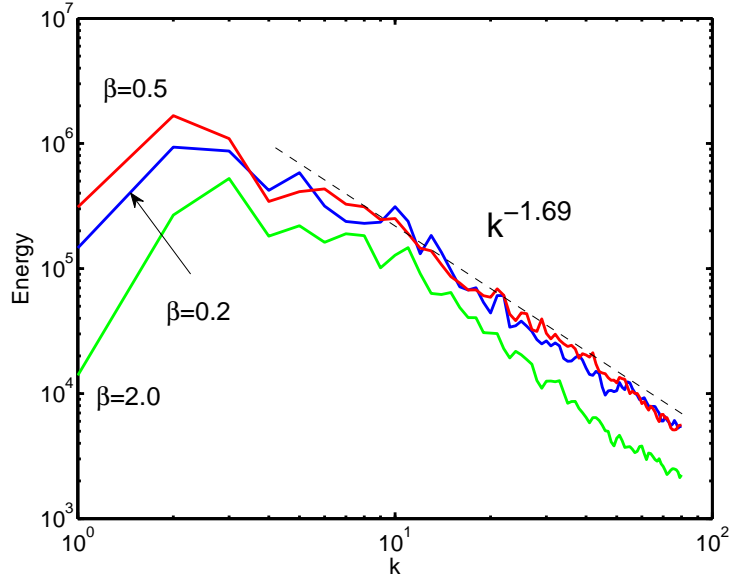}
\caption{\label{fig3} The 2D electron plasma turbulence in the
  presence of the density gradient (maintained by an external electric field) 
  exhibits a turbulent spectrum close to the Kolmogorov-like $k^{-5/3}$ scaling 
  over the inertial range. The effect of a  variation of the mean electric field
  is also shown in the  turbulent spectrum.}
\end{figure}

The power spectrum associated with the electron plasma turbulence
exhibits a spectral slope close to a Kolmogorov-like scaling $k^{-5/3}$, as
shown in \Fig{fig3}. This is indicative of the eddy interaction being
the dominant process in the spectral cascades of the inertial range
turbulent energy despite the presence of the convection of the density 
gradient by the mean electric field, and is consistent with the Kolmogorov-like 
phenomenology.

\begin{figure}
\includegraphics[width=12cm]{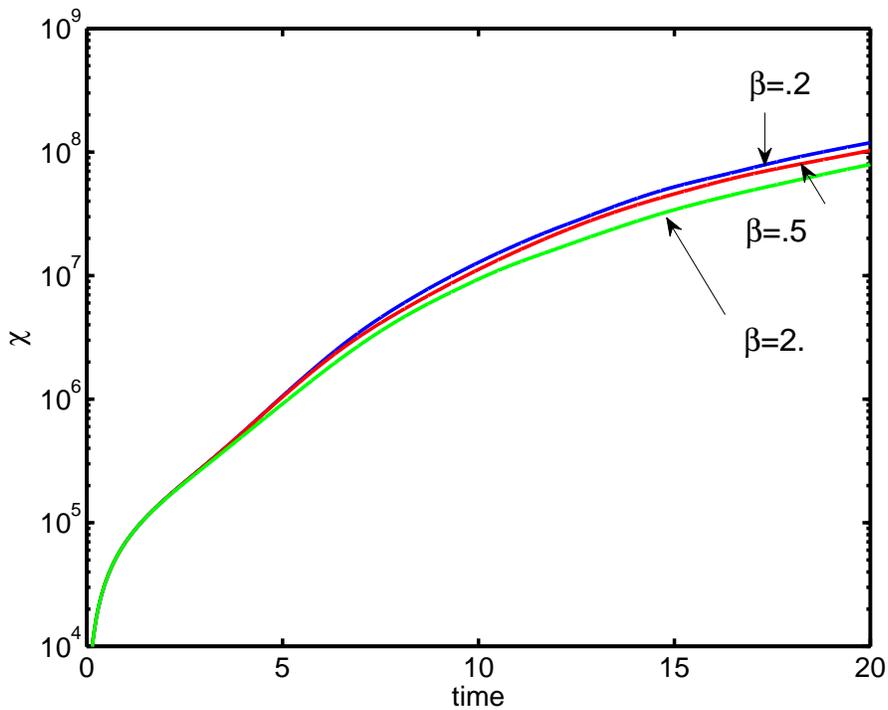}
\caption{\label{fig4} Evolution of the effective diffusion coefficient
  is studied in the presence of the density gradient in  a nonuniform 2D
  electron magnetoplasma. Increasing strength of the inhomogeneity leads 
  to a reduction in the net electron transport across the magnetic field
  lines. This result is also consistent with the experimental observations.}
\end{figure}

\section{Turbulence Transport}

The cross-field electron transport is triggered primarily by the mutual 
collisions of the electrons in the presence of an ensemble of coherent 
structures. Predominantly, it is the successive collisions amongst the
electrons that can potentially lead to a net outward drift 
across the external magnetic field. However, the presence of an
external orthogonal electric field is believed to improve the
stability of 2D electron plasma. Motivated by this observation, we
investigate here the transport of electrons that is influenced by
turbulent evolution of the potential and vorticity in 2D electron
plasma.  We follow the evolution of the effective diffusion by increasing
the magnitude of externally imposed mean electric field.  An effective
electron diffusion coefficient can be calculated from $ D_{eff} =
\int_0^\infty \langle {\bf V}_\perp({\bf r},t) \cdot {\bf V}_\perp
({\bf r},t+ t^\prime) \rangle dt^\prime$, where the angular
bracket represents spatial averages. The perpendicular component of
the electron fluid velocities in 2D electron plasma is ${\bf V}_\perp
\approx \hat {\bf z} \times \nabla \phi$.  Since the 2D EDM turbulence is
confined in a plane orthogonal to the ambient magnetic field in $\hat
{\bf z}$ direction, the effective cross-field diffusion coefficient,
$D_{eff}$, essentially describes the diffusion processes associated
with the transverse motion of the electrons in $x-y$ plane.  We compute
$D_{eff}$ in our simulations to measure the turbulent transport that
is associated with the intermittent electron plasma wave turbulence. It is
observed that the effective cross-field transport is lower, when the
field perturbations are Gaussian.  On the other hand, the cross-field
diffusion increases rapidly with the eventual formation of longer
length-scale structures. This is shown in \Fig{fig4}, which exhibits 
a dependence of $D_{eff}$ with time.  Thus,
an enhanced cross-field transport level results primarily due to an
emergence of large-scale coherent structures in a nonuniform magnetized
plasma. The enhanced cross-field diffusion coefficient observed in our
simulations is, therefore, consistent with the generation of large-scale flows. 

The most remarkable point to emerge from our simulations is the
reduction of the electron transport across the confining magnetic
field when the strength of the applied or external electric field is
increased. We have reported three different cases for the evolution of the
diffusion coefficient corresponding to $\beta=0.2, 0.5$ and $2.0$ in
\Fig{fig4}. It is clear from \Fig{fig4} that the effective transport 
associated with the electron plasma waves is much better (see the lower curve 
in \Fig{fig4} for $\beta=2.0$) for a higher strength of the applied or external 
electric field. Our simulation results are further consistent with the
experimental observation of Ref. \cite{driscol}.

\section{Discussion}

In this paper, we have studied the turbulent properties of nonlinearly
interacting low-frequency EDMs in a nonuniform magnetoplasma by using 
computer simulations. Specifically, we have numerically investigated 
the modified Navier-Stokes equation (2) and have studied spectral cascades 
and cross-field electron transport produced by nonlinearly saturated 
turbulent modes in such a plasma system. It is found that self-organized 
coherent structures emerge due to nonlinear interactions between different 
scale sizes 2D fluctuations. The final state is thus intermittent and exhibits 
a dual cascade in which a forward cascade of the electron vorticity coexists 
with an inverse cascade of the potential fluctuation. The inertial power spectrum 
is close to the Kolmogorov-like 5/3 scaling, whereas the transport is dominated 
by large scale vortical structures. We find that the stability of the electron plasma
column confined by a straight magnetic field can be improved by virtue of an 
external electric field maintaining the density inhomogeneity, a result that 
is also consistent with experimental observations \cite{driscol}.



\end{document}